\pdfoutput=1
\documentclass[a4paper,11pt]{article}

\usepackage{amsmath}
\usepackage{amssymb}
\usepackage{graphicx}
\graphicspath{{figures/}} 
\usepackage{xfrac}

\usepackage{braket}
\usepackage[squaren]{SIunits}

\usepackage[utf8]{inputenc}
\usepackage{todonotes}

\usepackage[square,numbers,comma,sort&compress]{natbib}
\usepackage[colorlinks=true,citecolor=blue,linkcolor=purple,urlcolor=purple]{hyperref}
\usepackage{geometry}
\usepackage{cleveref}
\usepackage{booktabs}
\usepackage{slashed}

\newcommand{\Tr}{\mathrm{Tr}}
\newcommand{\D}{\ensuremath{\,\text{d}}}

\newcommand{\imag}{\ensuremath{\text{i}}}

\newcommand{\alphaS}{\ensuremath{\alpha_\text{s}}}
\newcommand{\gstrong}{\ensuremath{g_\text{s}}}

\newcommand{\nf}{\ensuremath{n_\text{F}}}
\newcommand{\TR}{\ensuremath{T_\text{R}}}

\newcommand{\NC}{\ensuremath{N_\text{C}}}
\newcommand{\tev}{\ensuremath{\tera e\volt}}
\newcommand{\gev}{\ensuremath{\giga e\volt}}
\newcommand{\hc}{\mathrm{h.c.}\,}

\def\varabstract{ }
\def\varkeywords{ }
\def\vararxivnumber{ }
\def\vartitle{ }
\def\varsubtitle{ }
\renewcommand{\title}[1]{\gdef\vartitle{#1}}

\renewcommand{\abstract}[1]{\gdef\varabstract{#1}}
\newcommand{\keywords}[1]{\gdef\varkeywords{#1}}

\newtoks\authtoks
\renewcommand{\author}[2][]{%
	\authtoks=\expandafter{\the\authtoks#2$^{#1}$\ }%
}
\newtoks\affiltoks
\newcommand{\affiliation}[2][]{%
    \affiltoks=\expandafter{\the\affiltoks{\item[$^{#1}$]#2}}%
}
\newtoks\emailtoks\newcounter{emailcounter}%
\newcommand{\emailAdd}[1]{%
\ifnum\theemailcounter>0\emailtoks=\expandafter{\the\emailtoks, \typeemail{#1}}%
\else\emailtoks=\expandafter{\typeemail{#1}}%
\fi
\stepcounter{emailcounter}}
\newcommand{\typeemail}[1]{\href{mailto:#1}{\tt #1}}

\renewcommand\maketitle{
	\newgeometry{margin=2cm}
	\pagestyle{empty}\setcounter{page}{0}
	{\huge\flushleft\sffamily\bfseries\vartitle\\\Large\varsubtitle\par}
\vskip6ex
{\large\bfseries\raggedright\sffamily\the\authtoks\par}
\vskip2ex
\begin{list}{}{%
\setlength{\leftmargin}{0.28cm}%
\setlength{\labelsep}{0pt}%
\setlength{\itemsep}{-3pt}%
\setlength{\topsep}{-\parskip}}
\itshape\small%
\the\affiltoks
\end{list}
\vskip2ex
\noindent\hspace{0.28cm}\begin{minipage}[l]{.9\textwidth}
\begin{flushleft}
\textit{E-mail:} \the\emailtoks
\end{flushleft}
\end{minipage}
\vskip5ex
\noindent{\renewcommand\baselinestretch{.9}\textsc{Abstract:}}\ \varabstract
\vskip5ex
\if!\varkeywords!\else\noindent{\textsc{Keywords:}}\ \varkeywords \vskip2ex\fi
\if!\vararxivnumber!\else\noindent{\textsc{ArXiv ePrint:}} \href{http://arxiv.org/abs/\vararxivnumber}{\vararxivnumber}\vskip2ex\fi

\newpage
\restoregeometry
\pagestyle{plain}
\setcounter{footnote}{0}
}

\usepackage[square,numbers,comma,sort&compress]{natbib}

\definecolor{MS}{rgb}{0,0,1}

\usepackage{ifthen}

	\newcommand{\barlimc}[7]{
  \pgfmathparse{\mypos+0.3}
  \edef\mypos{\pgfmathresult}
		\node[left,scale=0.6] at (0,\mypos) {#1};
		\pgfmathparse{#3 > 5 ? 1 : 0}
		\ifthenelse{\pgfmathresult=1}{
			\fill[#2] ($(0,\mypos)+(0,-0.1)$) rectangle +(5,0.2);
			\fill[white] ($(0,\mypos)+(3.5,-0.1)$) rectangle +(0.3,0.2);
			\draw[decoration={zigzag},decorate,#2,very thick] (3.4,\mypos) to +(0.5,0);
			\node[left,scale=0.6] at (5,\mypos) {#3};
			}{
			\fill[#2] ($(0,\mypos)+(0,-0.1)$) rectangle +(#3,0.2);
			\node[left,scale=0.6] at (#3,\mypos) {#3};
		}
		\fill[#4] ($(0,\mypos)+(0,-0.1)$) rectangle +(#5,0.2);
		\node[left,scale=0.6] at (#5,\mypos) {#5};
		\fill[#6] ($(0,\mypos)+(0,-0.1)$) rectangle +(#7,0.2);
		\pgfmathparse{#7 <0.3 ? 1 : 0}
		\ifthenelse{\pgfmathresult=1}{
			\node[right,scale=0.6] at (0,\mypos) {#7};
		}{
		\node[left,scale=0.6] at (#7,\mypos) {#7};
	}
}

\title{On long-lived electroweak-singlet colored scalars}
\author[1]{Christian T Preuss}\emailAdd{Christian.Preuss@monash.edu}
\author[1]{and German Valencia}\emailAdd{German.Valencia@monash.edu}
\affiliation[1]{School of Physics and Astronomy, Monash University, Wellington Road, Clayton, VIC-3800, Australia}

\abstract{
There has been much recent interest in long-lived massive particles at the LHC, understood as those with lifetimes between tens of micrometers and several meters. In this context we consider the possibility of long-lived electroweak singlet scalars charged under color $\mathrm{SU}(3)$ with masses near a TeV. The shortest lifetime of interest is already longer than typical hadronization scales.  These exotic new particles would therefore appear as color singlet bound states of the new scalars with quarks and gluons and it is their color charge that prevents them from decaying. In particular we consider color representations consistent with maintaining asymptotic freedom, those with dimensionality $d_R \leq 15$. We find that only the octets can decay, and they do so into multi-jet final states through the two-gluon channel. The other representations are stable and form fractionally charged color singlets, with the decuplet being the only one that can form electrically neutral color singlets.
}

\keywords{long-lived colored scalars, fractionally charged}

\begin{document}

\maketitle

{\hypersetup{linkcolor=black}
  \tableofcontents}

\newpage

\section{Introduction}

Recently there has been renewed interest in the study of long-lived particles at the LHC, what is dubbed the lifetime frontier. In the context of LHC studies the interesting lifetime range lies between tens of micrometers and  several meters. Recent reviews that highlight different theoretical scenarios that may give rise to such long-lived particles and the experimental strategies to search for them are, for example, Refs.~\cite{Eberhardt:2021ebh,Alimena:2019zri,Curtin:2018mvb}.

One intriguing possibility consists of long-lived scalars charged under the color group. Due to color confinement, such particles would hadronize before decaying (if they decay at all) and show up in the detectors as exotic hadrons. Examples of these kinds of particles which have been studied at length in the literature are the so called $R$-hadrons \cite{FARRAR1978575}.  In this case, gluinos or squarks are pair produced and are either long-lived or stable due to $R$-parity. Other scenarios have also been considered for the case of color triplets \cite{Barnard:2015rba,delaPuente:2015vja}. For the electroweak singlet scalars we consider here, it is their color charge that prevents them from decaying, as will be outlined below.

We present a bottom-up simplified model in which we extend the standard model (SM) with a set of electroweak singlet scalars in a representation $R$ of the color $\mathrm{SU}(3)$. These objects would then be copiously pair produced by gluon fusion at the LHC if their masses are below the order of a TeV. Being electroweak singlets, their exotic color structure can prevent them from decaying and color thus plays the role of $R$-parity in making these particles stable. Carrying a color charge, these hypothetical scalars would also combine with quarks and gluons to form color-singlets on typical hadronization time scales. Experimental searches would thus mimic the strategies used for $R$-hadrons.

We find that the requirement of maintaining asymptotic freedom restricts the possible $\mathrm{SU}(3)$ representations to $R = \boldsymbol{3},~\boldsymbol{6},~\boldsymbol{8},~\boldsymbol{10},~\boldsymbol{15},~\boldsymbol{15^\prime}$. With masses below $\sim 1$~TeV, their production cross-sections would be in the $\sim 0.1-1000$~pb range and they can potentially introduce significant corrections to the Higgs decay which then serves to constrain them. We find that the octet is the only one of these examples that could decay into two gluons, and for most of parameter space would be short-lived. The others are stable and would hadronize into fractionally charged color singlets with the exception of the color decuplet which can form an electrically neutral exotic hadron.

This manuscript is structured as follows. In \cref{sec:gluonCouplings}, we outline the extension of the SM towards including the new colored scalars based on arguments of conserving asymptotic freedom. \Cref{sec:scalarPotential} focusses on the scalar potential of the new particles. \Cref{sec:scalarDecayModes} then discusses all possible decay modes at leading-order (LO) and next-to-leading order (NLO). Before concluding in \cref{sec:conclusions}, we briefly discuss implications on hadrons created by such new colored scalars in \cref{sec:exoticHadrons}.

\section{Couplings to gluons}\label{sec:gluonCouplings}
We begin by considering those model-independent couplings that depend only on the color representation of the scalars and that determine their tree-level couplings to gluons. 
For a complex electroweak-singlet scalar field $S$ transforming with respect to an irreducible representation $R$ of $\mathrm{SU}(3)$ its interaction with the gluon field $A^\mu$ is dictated by the covariant derivative in the quadratic part of the Lagrangian,
\begin{equation}
	\mathcal L = \left[\left(\partial_\mu +\imag \gstrong A_\mu^A T_R^A \right) S \right]^\dagger \left(\partial^\mu +\imag \gstrong A^{A,\mu} T_R^A \right) S-m_S^2 S^\dagger S.
	\label{eq:quadL}
\end{equation}
The coupling strength is the usual strong coupling $\gstrong$ and in writing \cref{eq:quadL} we have omitted explicit color indices. For real representations of $\mathrm{SU}(3)$, the scalar fields will be real and an additional factor of $1/2$ is required. 

\subsection{Asymptotic freedom}
The $\mathrm{SU}(3)$ representations that we consider are first restricted by requiring the model to maintain asymptotic freedom at scales above the scalar masses. To this end, we recall the well known QCD one-loop beta function, to which vector, fermion, and scalar fields charged under the $\mathrm{SU}(3)$ contribute as,
\begin{equation}
	\beta_g = -\frac{\gstrong^3}{16\pi^2}\left(\frac{11}{3}t_2(V) - \frac{4}{3}\nf t_2(F) - \frac{1}{3}t_2(S)\right). \label{eq:QCDBeta}
\end{equation}
Here, $t_2(V)$, $t_2(F)$, and $t_2(S)$ denote the Dynkin index of the representation in which the different fields transform respectively. For the SM augmented by complex scalars in representations with dimension up to 15, \cref{eq:QCDBeta} takes the form
\begin{equation}
	\beta_g =- \frac{\gstrong^3}{16\pi^2}\left(11 - \frac{2}{3}\nf - \frac{1}{6}n_3 - \frac{5}{6} n_6 - n_8 - \frac{5}{2}n_{10} - \frac{10}{3}n_{15} - \frac{35}{6} n_{15'}\right). 
\end{equation}
In what follows we will only allow one scalar multiplet at a time, and this implies that 
asymptotic freedom holds as long as\footnote{Similar considerations have been used for fermions in higher color representations  \cite{Marciano:1980zf,Chivukula:1990di}.}
\begin{equation}
	t_2(S) < 21.
\label{eq:afcon}
\end{equation}
All possible multiplets satisfying \cref{eq:afcon} are collected in \cref{tab:DynkinIndices} and their corresponding pair-production cross-section at the LHC is plotted in \cref{f:sigma}. The lowest-dimensional representation to fail \cref{eq:afcon} is $(3,1)$ for which $t_2(\mathbf{24})=25$.
\begin{table}[h]
	\centering
	\caption{Dynkin indices of all representations of colored scalars allowed by asymptotic freedom.}
	\label{tab:DynkinIndices}
	\begin{tabular}{ccc}\toprule
		\textbf{Label} & \textbf{Representation} & \textbf{Dynkin Index $t_2$}  \\ \midrule
		$(0,0)$ & $\boldsymbol{1}$ & $0$ \\
		$(1,0)$ & $\boldsymbol{3}$ & $1/2$ \\
		$(2,0)$ & $\boldsymbol{6}$ & $5/2$ \\
		$(1,1)$ & $\boldsymbol{8}$ & $3$ \\
		$(3,0)$ & $\boldsymbol{10}$ & $15/2$ \\
		$(2,1)$ & $\boldsymbol{15}$ & $10$ \\
		$(4,0)$ & $\boldsymbol{15}^\prime$ & $35/2$ \\ \bottomrule
	\end{tabular}
\end{table}

\subsection{Production via gluon fusion at the LHC}
\begin{figure}[t]
	\centering
	\includegraphics[width=0.9\textwidth]{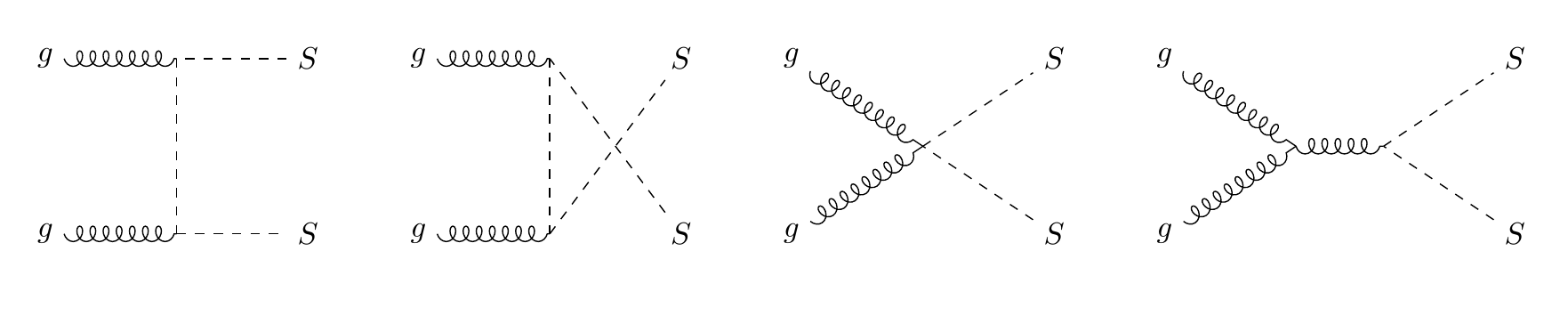}
	\caption{Leading-order diagrams for scalar pair production from gluon fusion.}
	\label{f:dia}
\end{figure}

\Cref{eq:quadL} determines the scalar pair-production cross-section from gluon fusion 
which proceeds at LO through the diagrams in \cref{f:dia}.
For complex scalars it is given in terms of the quadratic Casimir of the representation, $c_2(R)$, by~\cite{Manohar:2006ga}
\begin{align}
	\frac{\D \sigma}{\D t} & = \frac{2\pi\alphaS^2}{s^2} \frac{c_2(R){\dim}~R}{({\dim}~A)^2} \left[\left(c_2(R)-\frac{1}{4}c_2(A)\right)+\frac{c_2(A)}{4}\frac{(u-t)^2}{s^2}\right] \nonumber \\
	&\qquad \times \left[1+\frac{2m_S^2t}{(m_S^2-t)^2}+\frac{2m_S^2u}{(m_S^2-u)^2}+\frac{4m_S^4}{(m_S^2-t)(m_S^2-u)}\right].
\label{eq:sigma}
\end{align}
The terms depending on the Casimir of the adjoint representation, $c_2(A)$, originate from the s-channel diagram and $s,t,u$ are the usual Mandelstam variables. For the case of real scalar fields, \cref{eq:sigma}  must be multiplied by an additional factor of $1/2$. Many detailed phenomenological studies for octets, particularly electroweak doublets, exist in the literature \cite{Krasnikov:1995kn,Manohar:2006ga,Gresham:2007ri,Han:2010rf,Hayreter:2017wra,Hayreter:2018ybt,Miralles:2019uzg} and we have checked that the corresponding results agree with ours. In particular, \cref{eq:sigma} is half as large as the corresponding cross-section for squark pair production from gluon fusion for which there are two complex scalar doublets \cite{PhysRevD.31.1581}.

\begin{figure}[t]
	\centering
	\includegraphics[width=10cm]{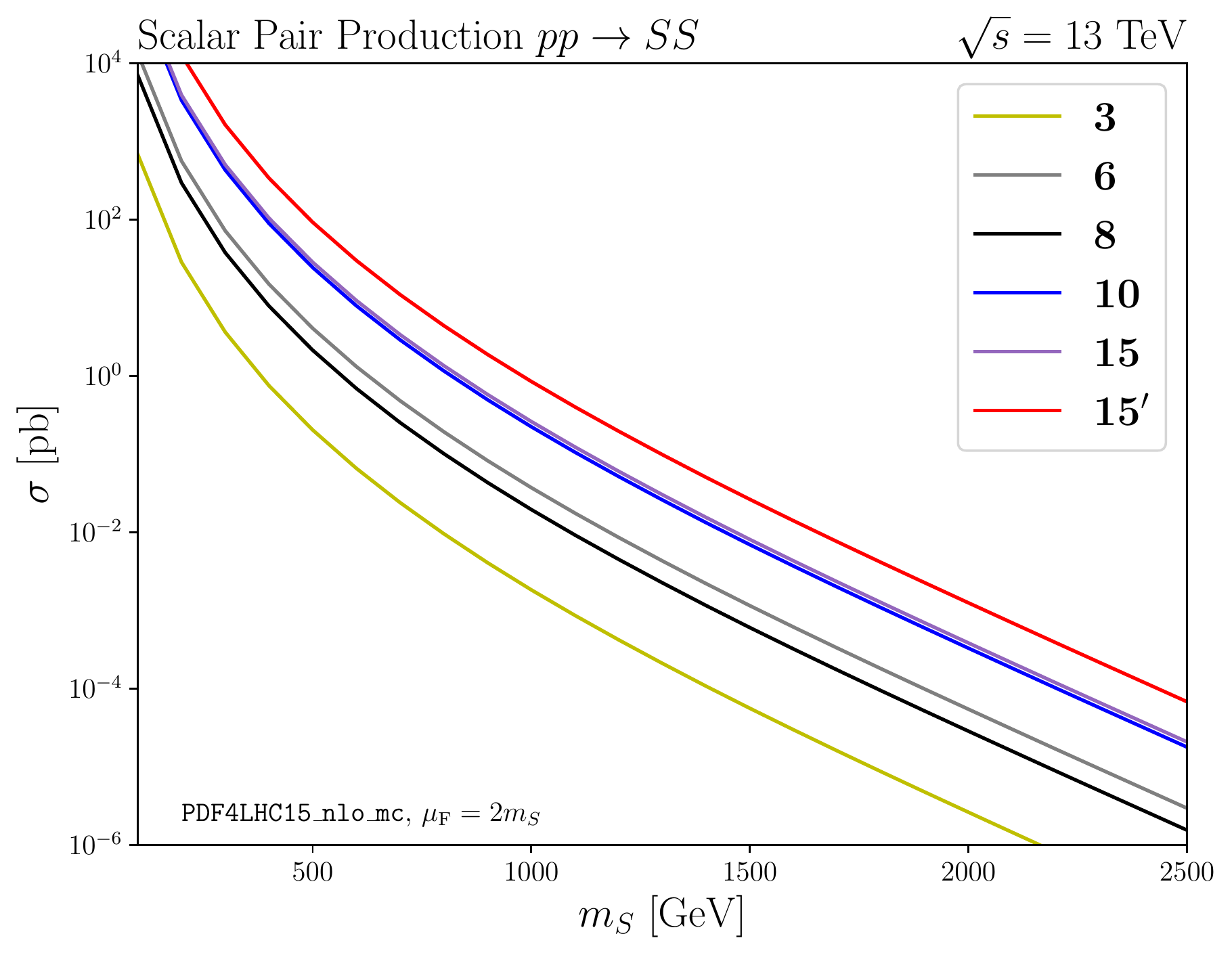}
	\caption{Leading-order scalar pair production cross-section $pp\to SS$ at the LHC with $\sqrt{s}=13~\tev$ for complex scalars in $\mathrm{SU}(3)$ representations $\mathbf{3}$, $\mathbf{6}$, $\mathbf{10}$, $\mathbf{15}$, $\mathbf{15^\prime}$ and real scalars in the adjoint representation $\mathbf{8}$.}
	\label{f:sigma}
\end{figure}

In \cref{f:sigma}, we present the leading-order $pp\to SS$ cross-section for all representations satisfying \cref{eq:afcon} at the LHC with $\sqrt{s}=13~\tev$ using the \texttt{PDF4LHC15\_nlo\_mc} parton distribution function set \cite{Butterworth:2015oua} with factorization scale $\mu_\mathrm{F}=2m_S$, accessed via the \texttt{ManeParse} package \cite{Clark:2016jgm} and \texttt{LHAPDF6} \cite{Buckley:2014ana}. 
Calculations of these cross-sections beyond leading order exist for the triplets (squarks) \cite{Beenakker:1996ch} where NLO enhancements over the LO are found to be described approximately by modest $K$-factors near 1.3 at $\sqrt{s}=14~\tev$ with weak dependence on the squark mass. Next-to-leading order calculations for scalar octets (sgluons) exist in the literature \cite{GoncalvesNetto:2012nt} as well, and a $K$-factor in the range (1.5-1.8) is found at $\sqrt{s}=14~\tev$, with the variation stemming from the sgluon mass. Leading-order color sextet (diquark) production at the LHC has been studied previously in \cite{Richardson:2011df}.

\section{The scalar potential}\label{sec:scalarPotential}
The renormalizable interactions of colored scalars are governed by the most general scalar potential up to dimension four. For electroweak singlets, the terms
\begin{equation}
    V_R = \kappa_R H^\dagger H S^\dagger S + \mu_R (S^\dagger S)^2 \, ,
    \label{eq:vrgen}
\end{equation}
occur for all $\mathrm{SU}(3)$ representations. There are also further representation-dependent terms, which will be detailed below for specific cases. 

Triple-scalar couplings are allowed when $R\otimes R\otimes R\  \supset \boldsymbol{1}$, and this is possible for all the representations we consider, as can be seen from \cref{prods34}. However, some of these couplings are completely antisymmetric under scalar exchanges and therefore vanish when the scalars carry no other quantum numbers beyond color. 
Completely symmetric triple-scalar couplings occur except for the $\boldsymbol{3}$ and $\boldsymbol{10}$ as shown below. We denote symmetric (anti-symmetric) triple couplings by $\lambda_R$ ($\tilde\lambda_R$). The triple-scalar couplings allowed under these conditions have dimensions of mass but they are not related to the usual Higgs vacuum expectation value. These presumably originate from the mechanism that gives mass to the scalars.

Model-dependent quartic-scalar couplings occur if $R\otimes R\otimes R\otimes R\  \supset \boldsymbol{1}$ and this occurs for $R = \boldsymbol{8}$, $\boldsymbol{10}$. In addition, for $R = \boldsymbol{10}$ there is a quartic coupling involving three scalar particles and one anti-particle.

In the following, fundamental indices of the color group are denoted by small Latin letters $i,j,k,\ldots$, while adjoint indices are denoted by capital Latin letters $A,B,C, \ldots$. The potentials for  $R=\boldsymbol{3},~\boldsymbol{6}$, motivated by minimal flavor violation, have been discussed in \cite{Arnold:2009ay} albeit with non-zero $U(1)_Y$ charge.

\paragraph{Triplets} It is possible to construct a singlet from three triplets but not from four so the most general potential is
\begin{equation}
    V_{\boldsymbol{3}} = m_S^2 S^\dagger_i S^i +\kappa_{\boldsymbol 3} H^\dagger H S^\dagger_i S^i +\mu_{\boldsymbol 3} (S^\dagger_i S^i)^2 + (\tilde\lambda_{\boldsymbol 3} \varepsilon_{ijk} S^i S^j S^k + \hc ).
\end{equation}
Note that the $\tilde\lambda_{\boldsymbol 3}$ term is completely antisymmetric and therefore vanishes for identical scalars. 

\paragraph{Sextets}
For scalar sextets it is once again possible to construct a singlet with three fields but not four, so the most general potential is given by:
\begin{equation}
    V_{\boldsymbol{6}} =m_S^2 S^\dagger_{ij} S^{ij} +\kappa_{\boldsymbol 6} H^\dagger H S^\dagger_{ij} S^{ij} +\mu_{\boldsymbol 6a} (S^\dagger_{ij} S^{ij})^2 +\mu_{\boldsymbol 6b}    S^\dagger_{ik} S^\dagger_{j\ell} S^{ij} S^{k\ell}
    + (\lambda_{\boldsymbol 6} \varepsilon_{ikm}\varepsilon_{j\ell n} S^{ij} S^{k\ell} S^{mn}  + \hc).
\end{equation}
In this case the triple-scalar coupling  is symmetric under the exchange of any two sextets.

\paragraph{Octets} 
The octet is a real representation of $\mathrm{SU}(3)$ and, in the absence of  additional quantum numbers, we will use a real scalar field. As the octet plays a special role as the adjoint representation, we write the potential for this case using adjoint indices,\footnote{A real scalar in the model of \cite{Manohar:2006ga} is $S_R$ and the triple-scalar coupling would correspond to $\lambda_{\boldsymbol{8}}=v(\lambda_4+\lambda_5)/12$ in that model.} 
\begin{equation}
    V_{\boldsymbol{8}} = m_S^2 S^A S^A  + \kappa_{\boldsymbol 8} H^\dagger H S^A S^A +\mu_{\boldsymbol 8} (S^A S^A)^2 + \lambda_{\boldsymbol 8} d_{ABC} S^A S^B S^C 
\end{equation}
Another quartic term found in the literature, $\sim{\rm Tr}(T^AT^BT^CT^D) S^A S^B S^C S^D$ \cite{Krasnikov:1995kn,Dobrescu:2011aa},  reduces to $\mu_8$ when the scalars do not carry other quantum numbers.

\paragraph{Decuplets} For scalar decuplets, the most general potential is given by:
\begin{align}
    V_{\boldsymbol{10}} &= m_S^2 S^\dagger_{ijk} S^{ijk} +\kappa_{\boldsymbol{10}} H^\dagger H S^\dagger_{ijk} S^{ijk} +\mu_{\boldsymbol{10}a} (S^\dagger_{ijk} S^{ijk})^2 
    +\mu_{\boldsymbol{10}b} S^\dagger_{ijk} S^\dagger_{\ell m n}S^{ijm}S^{\ell n k} \nonumber \\
    &+ \left( \tilde\lambda_{\boldsymbol{10}} \varepsilon_{i\ell o}\varepsilon_{jmp}\varepsilon_{knq} S^{ijk} S^{\ell mn} S^{opq}+\omega_{\boldsymbol{10}} \varepsilon_{i\ell o}\varepsilon_{jmr}\varepsilon_{kps} \varepsilon_{nqt}S^{ijk} S^{\ell mn} S^{opq} S^{rst}\right. \nonumber \\
    &+ \left.\rho_{\boldsymbol{10}} S^\dagger_{knq}~\varepsilon_{i\ell o}\varepsilon_{jmp}S^{ijk} S^{\ell mn} S^{opq} + \hc\right)
\end{align}
This color structure allows an anti-symmetric triple-scalar vertex, a second quartic vertex with two particles and two anti-particles, a quartic vertex with four particles and a quartic vertex with three particles and one anti-particle.

\paragraph{$(2,1)$-Quindecuplets}
For scalars transforming in the $\mathbf{15}$, we find the potential
\begin{align}
    V_{\boldsymbol{15}} &= m_S^2 S^{\dagger k}_{ij} S^{ij}_{k} +\kappa_{\boldsymbol{15}} H^\dagger H  S^{\dagger k}_{ij} S^{ij}_{k} +\mu_{\boldsymbol{15}a} ( S^{\dagger k}_{ij} S^{ij}_{k})^2
    +\mu_{\boldsymbol{15}b} (S^{\dagger k}_{ij} S^{\dagger \ell}_{mn} S^{im}_{k}S^{jn}_{\ell})   +\mu_{\boldsymbol{15}c} (S^{\dagger k}_{ij} S^{\dagger \ell}_{mn} S^{im}_{\ell}S^{jn}_{k}) \nonumber \\ 
    &+
    \left( \lambda_{\boldsymbol{15}}\varepsilon_{\ell mn}S^{i\ell}_{j} S^{jm}_{k} S^{kn}_{i}+ \tilde\lambda_{\boldsymbol{15}}\varepsilon_{ijk}\varepsilon_{\ell mn}\varepsilon^{o p q}S^{i\ell}_{o} S^{jm}_{p} S^{kn}_{q}  + \hc\right)
\end{align}
The $(2,1)$ representation allows a symmetric triple-scalar vertex, the term with coupling $\lambda_{\boldsymbol{15}}$; as well as an antisymmetric one with coupling $\tilde\lambda_{\boldsymbol{15}}$. 
It does not allow quartic terms with three particles and one antiparticle nor ones with four particles.

\paragraph{$(4,0)$-Quindecuplets}
For scalars transforming in the $\mathbf{15}^\prime$, we find the potential
\begin{align}
    V_{\boldsymbol{15^\prime}} &= m_S^2 S^\dagger_{ijk\ell} S^{ijk\ell} +\kappa_{\boldsymbol{15^\prime}} H^\dagger H S^\dagger_{ijk\ell} S^{ijk\ell} +\mu_{\boldsymbol{15^\prime}a} (S^\dagger_{ijk\ell} S^{ijk\ell})^2 +\mu_{\boldsymbol{15^\prime}b} (S^\dagger_{ijk\ell} S^\dagger_{mnop} S^{ijmn}S^{k\ell op}) \nonumber \\
    &+ \mu_{\boldsymbol{15^\prime}c} (S^\dagger_{ijk\ell} S^\dagger_{mnop} S^{ijkm}S^{\ell nop})
   + \left( \lambda_{\boldsymbol{15^\prime}} \varepsilon_{i\ell o}\varepsilon_{jmp}\varepsilon_{knq} \varepsilon_{rst}S^{ijkr} S^{\ell mns} S^{opqt}+ \hc\right)
\end{align}
The $(4,0)$ representation permits a symmetric triple-scalar coupling but no quartic couplings.

We want to close this section by elaborating upon possible Yukawa couplings of the colored scalars.
Since we do not consider any Standard-Model extensions beyond the colored scalars, we only have the SM fermion content at our disposal, meaning that all fermions must either be color singlets or triplets. 
In effect, only scalar singlets, triplets, sextets, and octets can have Yukawa couplings. Moreover, if the scalars are $\mathrm{SU}(2)_L$ singlets and do not carry electric charge, the only possible Yukawa coupling is to right handed neutrinos. 
Insisting on $\mathrm{SU}(2)_L$ singlets but allowing for $Y=Q\neq 0$, the only possibilities with colored scalars are for triplets and sextets such as  \cite{Davies:1990sc}
\begin{align}
	V_{Y,\mathbf{3}} &\supset g_{\text{Y},\boldsymbol 3}^{e1} S^i \bar{d}_{i R} e^c_R
	+ g_{\text{Y},\boldsymbol 3}^{e2} S^i \bar{Q}_{i L} L^c_L
	+ g_{\text{Y},\boldsymbol 3}^{e3} S^i \bar{u}_{i R} e^c_R  \nonumber \\
	&\qquad +g_{\text{Y},\boldsymbol 3}^{q1} \varepsilon_{ijk} S^i \bar{Q}_L^c Q_L +g_{\text{Y},\boldsymbol 3}^{q2} \varepsilon_{ijk} S^i \bar{u}_R^c d_R
	+g_{\text{Y},\boldsymbol 3}^{q3} \varepsilon_{ijk} S^i \bar{u}_R^c u_R
	+g_{\text{Y},\boldsymbol 3}^{q4} \varepsilon_{ijk} S^i \bar{d}_R^c d_R + \hc \\
	V_{Y,\mathbf{6}} &\supset g_{\text{Y},\boldsymbol 6}^{q1} S^{ij} \bar{Q}_{iL} Q_{jL}^c
	+g_{\text{Y},\boldsymbol 6}^{q2} S^{ij} \bar{d}_{iR} u_{jR}^c
	+g_{\text{Y},\boldsymbol 6}^{q3} S^{ij} \bar{u}_{iR} u_{jR}^c
	+g_{\text{Y},\boldsymbol 6}^{q4} S^{ij} \bar{d}_{iR} d_{jR}^c +\hc
\label{eq:yukterms}
\end{align}
where we have omitted flavor indices.
All of these couplings require the scalars to carry electric charge, so are not of interest here.

\section{Scalar decay modes}\label{sec:scalarDecayModes}
We now examine possible decay modes at tree and one-loop level, and show that only the octet can decay. To conserve color, any scalar decay must include quarks, gluons or more scalars. 

\subsection{Tree-level decays}
Decays into standard-model fermion pairs, $S\to f\bar f$, could proceed at tree-level through Yukawa interactions. 
As the fermions in the SM transform under the color group as singlets or triplets, the $f\bar f$ state can only be an $\mathrm{SU}(3)$ 
$\boldsymbol{1}$, $\boldsymbol{3}$, $\boldsymbol{6}$, or $\boldsymbol{8}$. 
Requiring that the new scalars be electroweak singlets forbids any couplings of the form listed in \cref{eq:yukterms} so decays to fermion pairs are not possible.  If instead, we allow 
scalars with non-zero hyper-charge, these would decay into two or more jets for $\boldsymbol{3}$, $\boldsymbol{6}$, or $\boldsymbol{8}$ and small Yukawa couplings would be required to ensure longevity.

Since we consider only one (degenerate) multiplet at a time, the scalars cannot decay into each other either. 

\subsection{One-loop decays}
\begin{figure}[t]
	\centering
	\includegraphics[width=0.9\textwidth]{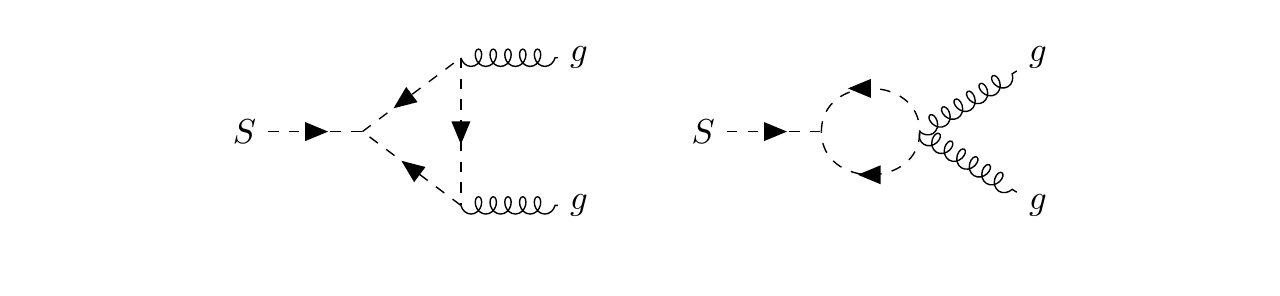}
	\caption{Lowest-order loop-induced diagrams for $S\to gg$. }
	\label{f:diasgg}
\end{figure}

The decay $S \to gg$ is in principle allowed as a loop-induced process. Without Yukawa couplings, however, it can only proceed via intermediate colored scalars as shown in \cref{f:diasgg}. This process then depends on the existence of appropriate triple-scalar couplings and can proceed only for scalars in the octet, $\mathbf{8}$, as we will show below.

Color conservation indicates that only those representations  that appear in the Clebsch-Gordan decomposition of the direct product of two adjoint representations can decay into two gluons,
\begin{equation}
	\mathbf{8} \otimes \mathbf{8} = \mathbf{1} \oplus _2 \mathbf{8} \oplus \mathbf{10} \oplus \overline{\mathbf{10}} \oplus \mathbf{27} 
\end{equation}
Of these, we only need to consider the adjoint $\mathbf{8}$ and (anti-)decuplet $\mathbf{10}$ ($\overline{\mathbf{10}}$) representations as per \cref{eq:afcon}.
It is straightforward to confirm that these are the only representations to consider even if a few additional gluons are added in the final state.

\begin{figure}[t]
	\centering
	\includegraphics[width=0.7\textwidth]{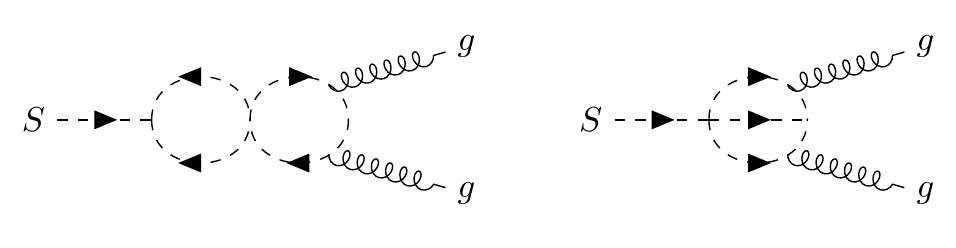}
	\caption{Generic two-loop diagrams for the $S_{\mathbf{10}}\to gg$ decay combining vertices with $\kappa_{\mathbf{10}}$ and $\mu^\prime_{\mathbf{10}}$. Arrows indicate particles/antiparticles (or color/anticolor representations). The gluons can be emitted from any of the scalar lines as shown or both at a time from a seagull vertex.}
	\label{f:dec2l}
\end{figure}

We begin by examining the case of the $\mathbf{10}$ (or $\overline{\mathbf{10}}$). The color flow (indicated by the arrows)  shows that for the left-hand diagram in \cref{f:diasgg} to be non-vanishing, either an $SSg$ or an $S^\dagger S^\dagger g$ vertex is required, while for the seagull diagram in \cref{f:diasgg} an $SSgg$ vertex is needed. 
However, gluons only appear in the quadratic part of the Lagrangian which couples a $\mathbf{10}$ with a $\overline{\mathbf{10}}$ due to the constraint of the Lagrangian being a color singlet.
Hence, neither of the one-loop diagrams appears for scalars transforming as (anti-)decuplets. This argument indicates that the diagrams in \cref{f:diasgg} only occur for real representations. 
We note that the decuplet also cannot decay into two gluons at the two-loop level when it carries no quantum numbers other than color. This is true, because the required triple-scalar vertex, cf.~\cref{f:dec2l} (the gluons can be attached to any scalar line), is antisymmetric and therefore vanishes for identical scalars.

Following the above reasoning, the only possibility left for $S \to gg$ to occur is for color octets where the answer at one-loop is already known. In adjoint indices it is given by \cite{Gresham:2007ri,Hayreter:2018ybt}
\begin{equation}
	{\cal L}_{Sgg} = \frac{9\alphaS}{8\pi}~\frac{\lambda_{\mathbf{8}}}{m_S^2}~I_s(1)~d_{ABC} G_{\mu\nu}^A G^{B\mu\nu} S^C,
\end{equation}
where the loop factor is $I_s(1)=\left(\frac{\pi^2}{9}-1\right)$ for the case where all intervening scalars are degenerate in mass, cf.~\cref{sec:loopfun}. The decay width is then 
\begin{equation}
	\Gamma(S\to gg) = \frac{9}{8\pi}\left(\frac{\alphaS}{8\pi}\right)^2~\frac{\lambda_{\mathbf{8}}^2}{m_S}~30~I_s(1)^2,
\end{equation}
with 30 being a color factor.

\subsection{Mixed decays to quarks and gluons}
Color conservation implies that the $S_{\mathbf{15}}$ can decay into states like $qg$ or $qgg$ and the $S_{\mathbf{15}^\prime}$ into  $qgg$. To proceed at dimension six (eight) these processes would  require  operators of the form ($\ell,q$ denote a generic lepton and quark respectively)
\begin{eqnarray}
{\cal O}\sim \bar{\ell}\sigma_{\mu\nu}q G^{\mu\nu} S_{\mathbf{15}},&&{\cal O}\sim \bar{\ell}q G^{\mu\nu}G_{\mu\nu} S_{\mathbf{15}^\prime},
\end{eqnarray}
where the gluons have to appear explicitly to satisfy the color requirement. The necessary fermion bilinears, however,  cannot couple to scalars with zero hyper-charge as we saw in our discussion of Yukawa couplings.

\section{Constraints from $H\to gg$}
\begin{figure}[t]
	\centering
	\includegraphics[width=0.9\textwidth]{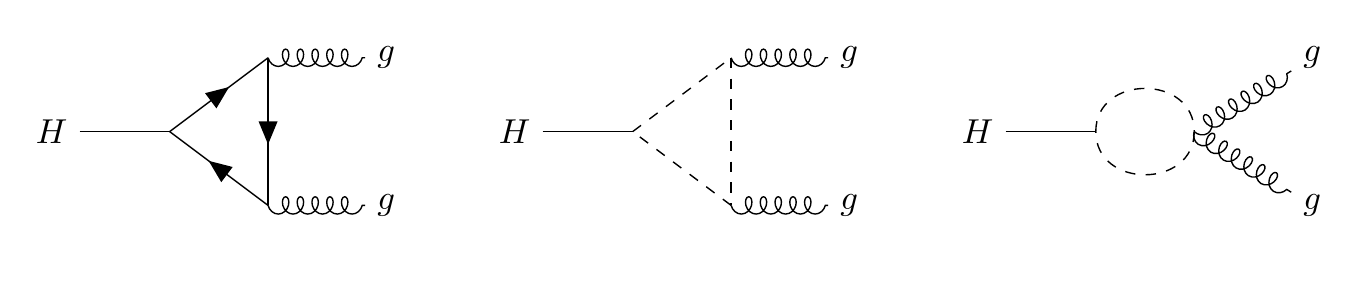}
	\caption{Lowest-order loop-induced diagrams for the decay $H\to gg$, including the standard-model contribution mediated by heavy quarks (\textit{left}) and new contributions mediated by colored scalars (\textit{middle} and \textit{right}).}
	\label{f:diahgg}
\end{figure}

For all representations we have discussed here, there is an additional new contribution to the Higgs decay $H\to gg$, and correspondingly to the Higgs boson production cross-section from gluon fusion, proceeding through the diagrams in \cref{f:diahgg}~\cite{He:2011ti,Dobrescu:2011aa}.
This contribution will be proportional to the $HSS$ coupling, which is contained in the term in the scalar potential proportional to $\kappa_R$ in \cref{eq:vrgen}.

\begin{figure}[t]
	\centering
	\includegraphics[width=0.6\textwidth]{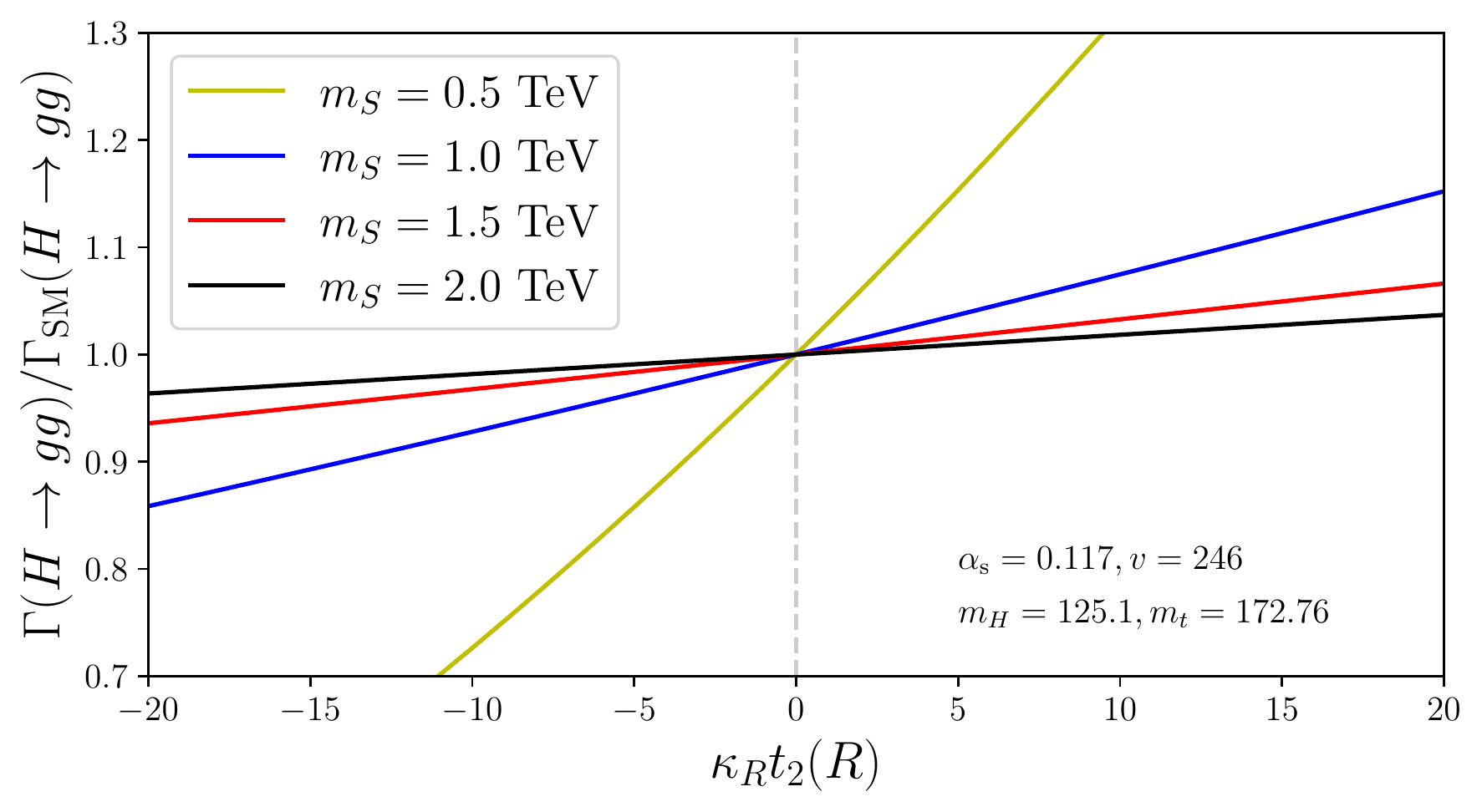}
	\caption{Corrections to the decay $H\to gg$ for different scalar masses and representations.}
	\label{f:hgg}
\end{figure}

After adding the SM top-quark loop contribution we find\footnote{In this notation, for example, the complex $S^\pm$ scalar field in \cite{Manohar:2006ga} would contribute with $\kappa_R=\frac{\lambda_1}{2}$ and $t_2(\mathbf{8})=3$.}
\begin{equation}
	\Gamma(H\to gg) = \frac{m_H^3}{8\pi v^2}\left(\frac{\alphaS}{\pi}\right)^2\left|I_q\left(\frac{m_t^2}{m_H^2}\right)+\kappa_R~\frac{v^2}{4m_H^2}t_2(R)~I_s\left(\frac{m_S^2}{m_H^2}\right)\right|^2
\end{equation}
with the known loop functions listed in \cref{sec:loopfun}. 
The corresponding corrections to the partial decay width of the Higgs $\Gamma(H\to gg)$ are illustrated in \cref{f:hgg}. From the figure we see that for $\vert \kappa_R \vert \lesssim 1$, and for all the representations satisfying \cref{eq:afcon} scalars with masses $m_S\gtrsim 1$~TeV have modest effects. The parameter space for lighter scalars, on the other hand, can be significantly constrained by Higgs decays.

\section{Exotic showers and hadrons}\label{sec:exoticHadrons}
As shown above, electroweak singlet scalars in color representations $R = \boldsymbol{3},~\boldsymbol{6},~\boldsymbol{10},~\boldsymbol{15},~\boldsymbol{15^\prime}$ cannot decay and are therefore long-lived. 
The radiation of additional gluons in form of bremsstrahlung is, however, possible for all the scalars we have considered here. 
After production, these scalars will therefore lose energy via a QCD showering process and, once a scale of the order of the hadronization scale $\Lambda_\mathrm{QCD} \approx 1~\gev$ is reached, hadronize by picking up quarks and gluons to form exotic color singlets. This situation is comparable to the one for so-called $R$-hadrons \cite{FARRAR1978575,Farrar:2010ps}, where exotic hadrons are formed from long-lived super-symmetric particles such as the stop or the gluino.
Irrespective of their origin, we will call such exotic hadrons $R$-hadrons in the following.

\begin{table}[h!]
	\centering
	\caption{List of possible $R$-hadrons for each long-lived representation. Here, $u$ and $d$ denote generic up- and down-type quarks, respectively.}
	\begin{tabular}{ll} \toprule
		Representation & $R$-hadrons \\\midrule
		$\boldsymbol{3}$ & $R_{S\bar{u}}^{-2/3}$, $R_{S\bar{d}}^{+1/3}$, $R_{Suu}^{+4/3}$, $R_{Sdd}^{-2/3}$ \\
		$\boldsymbol{6}$ & $R_{S\bar{u}\bar{u}}^{-4/3}$, $R_{S\bar{d}\bar{d}}^{+2/3}$, $R_{Sug}^{+2/3}$, $R_{Sdg}^{-1/3}$ \\
		$\boldsymbol{10}$ & $R_{Sgg}^{0}$ \\
		$\boldsymbol{15}$ & $R_{S\bar{u}g}^{-2/3}$, $R_{S\bar{d}g}^{+1/3}$ \\ 
		$\boldsymbol{15}'$ & $R_{S\bar{u}gg}^{-2/3}$, $R_{S\bar{d}gg}^{+1/3}$ \\ \bottomrule
	\end{tabular}
	\label{tab:rhadrons}
\end{table}
 
The simplest $R$-hadrons that can be formed from the long-lived scalars are collected in \cref{tab:rhadrons} with the corresponding Clebsch-Gordan decompositions shown in \cref{tab:rhadronsCG}. Except for the decuplet, all scalars would form fractionally charged hadrons. In addition, color singlets containing the $(4,0)$-quindecuplet can only be formed with a minimum of three additional partons.

Another intriguing possibility is the formation of quarkonium-like bound states following the production of 
scalar anti-scalar pairs. This was considered in \cite{Idilbi:2010rs} for color octets, and we will not pursue it further.

\subsection{Phenomenological modeling}
Multi-purpose event generators such as \textsc{Sherpa} \cite{Bothmann:2019yzt}, \textsc{Herwig} \cite{Bellm:2019zci}, and \textsc{Pythia} \cite{Sjostrand:2014zea} provide general tools for phenomenological studies of Standard Model phenomena as well as extensions to it. It would therefore be beneficial to implement our models considered above in such a framework to give access to a detailed modeling of not only the hard production process (as described by the leading-order cross-section in \cref{eq:sigma}), but also the modeling of logarithmically enhanced QCD bremsstrahlung and non-perturbative effects such as hadron fragmentation. 
In addition, Monte Carlo event generators provide the input for dedicated detector simulation programs such as \textsc{Geant4} \cite{GEANT4:2002zbu} or \textsc{Delphes} \cite{Ovyn:2009tx,deFavereau:2013fsa}, giving access to realistic particle-level analysis environments.

Monte Carlo event generators, however, routinely use the Les Houches standard to store color information \cite{Alwall:2006yp}, which only allows for two color tags per particle. (Strictly speaking it only allows for a color-anticolor combination, but this can be tweaked to enable color-color or anticolor-anticolor combinations, cf.~e.g.~\cite{Desai:2011su}.)
Implementations of models with colored particles in representations other than the (anti-)triplet or octet have therefore been limited mainly to sextets. An explicit ``color flow'' representation of the Standard-Model vertices has been presented in \cite{Kilian:2012pz} in the context of the \textsc{O'Mega} \cite{Ohl:2000hq} matrix element generator in \textsc{Whizard} \cite{Ohl:2000pr,Stienemeier:2021cse}. Sextet production including subsequent parton showering has been studied in \cite{Richardson:2011df}, while a general shower framework for heavy colored particles has been developed in \cite{Brooks:2019xso} and extended to showers off stops $\tilde{t}$ in \cite{Begic2019}.
As far as the non-perturbative modeling is concerned, both the \textsc{Herwig} and \textsc{Pythia} event generation frameworks provide dedicated modules for the hadronization of long-lived colored particles, in the form of $R$-hadrons, cf.~\cite{Desai:2011su,Desai:2021jsa} for a description of the model implemented in \textsc{Pythia}~8. A comparison of \textsc{Pythia}'s string and \textsc{Herwig}'s cluster fragmentation models for the case of $R$-hadrons was considered in \cite{Fairbairn:2006gg}.

A Monte Carlo implementation of the decuplet and quindecuplet models considered in this work necessitates the extension of the format in which color information is stored and interpreted in event generators. As the bare minimum, a third color tag has to be introduced and leading-order matrix elements projected onto a ``leading-color'', i.e.\ planar, decomposition in terms of these. It must be emphasized that while this may enable the parton-showering of such colored particles, it does not automatically include non-perturbative modeling of showered events. To this end, hadronization models require a more fundamental extension to utilize the additional color representations in the formation of color-singlet hadrons. 
Such an implementation is therefore beyond the scope of this work. We will instead obtain a first estimate of LHC bounds in the next section.

\subsection{A first estimate of LHC limits}
To estimate the limits that could be placed at LHC we can readily adapt some of the existing results for $R$-hadrons found in the literature, cf.\ e.g.\ \cite{Sirunyan:2017sbs,Aaboud:2019trc,Aad:2021yej}.

The ATLAS experiment, using $36.1~\femto\barn^{-1}$ of data at $\sqrt{s}=13~\tev$, rules out  at the 95\% c.l. long-lived sbottom and stop $R$-hadrons with masses below 1250~GeV and 1340~GeV respectively. The production cross-section for sbottom or stop pairs from gluon fusion\footnote{The leading-order gluon-fusion partonic cross-section for complex triplet scalars is the same as for squarks, but the one for octets is smaller than the one for gluinos, see for example  \cite{PhysRevD.31.1581,Beenakker:1996ch}.} is twice as large as the one shown in yellow for color triplets in \cref{f:sigma}. Differences in the lifetime are not important for these constraints, as the scalars considered here are stable and those in \cite{Aaboud:2019trc} are assumed to live long enough to reach the hadronic calorimeter (decay lengths of a few meters). For comparison, searches (with significantly lower integrated luminosity) assuming the particles to be stable find an upper bound for stop $R$-hadrons near $1~\tev$, cf.\ the respective CMS \cite{Khachatryan:2016sfv} and ATLAS \cite{Aaboud:2016uth} publications. The main difference would be in the constituents of the color-singlet exotic hadron. Whereas for sbottom and stop we could have neutral objects, for the color triplet considered here the exotic hadron would have a fractional electric charge, cf.~\cref{tab:rhadrons}. CMS searches looking for long-lived fractionally charged objects find similar constraints as the ones above, cf.\ \cite{CMS:2012xi,Khachatryan:2016sfv}. We thus expect bounds for exotic hadrons from color triplets to be around a TeV, slightly weaker than those for squarks, with stronger limits for the higher representations.

\begin{figure}[t]
\centering{\includegraphics[width=10cm]{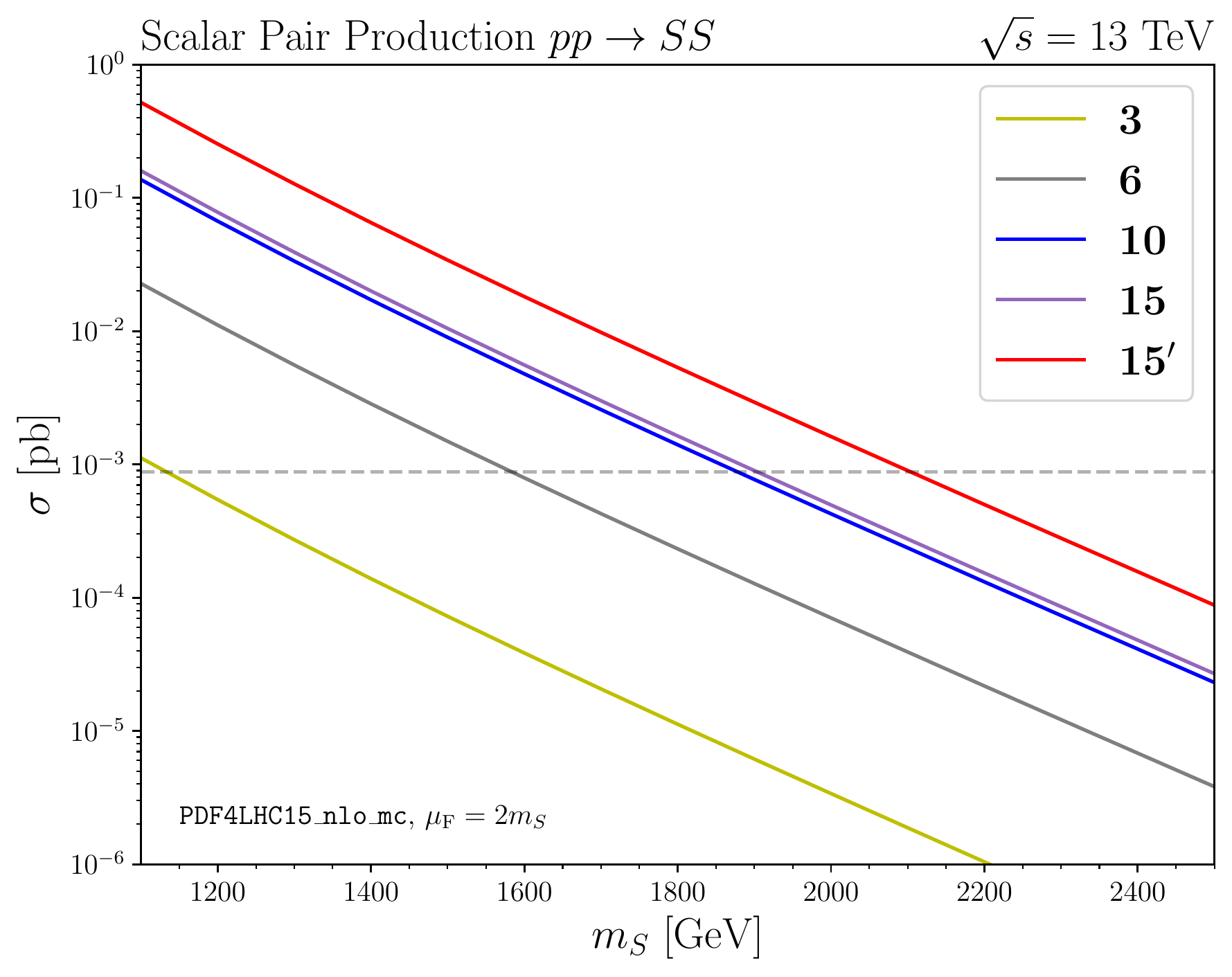}}
\caption{Pair-production cross-sections for all long-lived scalars including a $K$-factor of $1.3$, with the dashed line marking a first-estimate LHC bound.}
\label{f:bounds}
\end{figure}

To obtain a first estimate for the scalar-mass constraints, we compare our scalar pair-production cross-sections to the average cross-section corresponding to the ATLAS sbottom and stop $R$-hadron mass limits. In addition, we include a $K$-factor of $1.3$ for all representations. The results are shown in \cref{f:bounds}, where we have superimposed the first-estimate exclusion cross-section as a horizontal line. From this figure, the expected constraints based only on a scaling of the cross-section can be read off to range from $m_S = 1.1~\tev$ for color triplets to $m_S \approx 2.1~\tev$ for scalars in the $\mathbf{15^\prime}$.

As LHC limits for long-lived charged particles rely on measurements of ionization and as such do not directly apply to our neutral decuplet $R^0_{Sgg}$, some remarks are in order:
\begin{itemize}
\item First, the scalar decuplet can also hadronize with three quarks leading to a charged $R^+_{S\bar{u}\bar{u}\bar{d}}$ for example. The hadronization ratios into the different possible $R$-hadrons are not known, but are expected to be comparable. In addition,  collisions of an $R$-hadron with the detector material can change the electric charge through strong interaction reactions such as $R^0_{Sgg} + p\to R^+_{S\bar{u}\bar{u}\bar{d}} +{\rm pions}$. 
\item The neutral $R$-hadrons would also leave a hadronic shower in the hadronic calorimeters providing another handle for their detection.
\end{itemize}
It is interesting to note that one can also obtain an upper limit on the mass of neutral $R$-hadrons by requiring their annihilation cross-section be large enough to avoid over-production of relic dark matter density \cite{Busoni:2014gta,Albert:2017onk}. We can estimate the annihilation cross-section in a spectator model by adjusting the kinematic and color-average factors in \cref{eq:sigma} to find
\begin{equation}
        \langle\sigma(R^0_{Sgg}R^0_{Sgg}\to gg)v\rangle \approx \frac{\pi \alpha_s^2 c_2(R)}{8({\dim}~R )m_S^2}\left(4 c_2(R)-c_2(A)\right) \, .
\end{equation}
This result, in combination with $\langle\sigma v\rangle \gtrsim 4\times 10^{-9}~\gev^{-2}$ \cite{Busoni:2014gta}, suggests that $m_{R^0_{Sgg}}\lesssim 3.2~\tev$. There is no lower bound in this case since there could be many other contributions to the relic density.

\section{Summary and conclusions}\label{sec:conclusions}
In this manuscript, we have discussed scalar extensions to the Standard Model, transforming non-trivially under the color group and as singlets under the electroweak gauge group. Based on the requirement to leave asymptotic freedom intact, the maximal dimension of the color representation is 15. Requiring the new scalars to have zero hypercharge, the new particles are stable, with only octets decaying to gluons at the one-loop level. 

We have discussed how long-lived colored scalars will appear as exotic hadrons in detector signals and obtained a first estimate of mass bounds at the LHC. More detailed analyses can be obtained by implementing models of higher color representations in multi-purpose event generators and we have outlined such implementations, including a discussion of current technical restrictions.
The implementation in multi-purpose event generators is beyond the scope of the current work, but will be important to obtain mass constraints from  collider experiments, as it facilitates the modeling of QCD as well as EW/QED bremsstrahlung, hadronization as well as other non-perturbative corrections, and subsequent detector simulation.

\section*{Acknowledgements}
We would like to thank Ulrik Egede, Xiao-Gang He, Peter Skands  and Ray Volkas for helpful discussions.
CTP is supported by the Monash Graduate Scholarship, the Monash International Postgraduate Research Scholarship, and the J.~L.~William Scholarship.

\appendix

\section{Some $\mathrm{SU}(3)$ relations and notation}
For a general representation $R$, the Dynkin index is given by the trace of the generators $T_R$ in the representation,
\begin{equation}
	\Tr(T_R^A T_R^B) = t_2(R) \delta^{AB},
\end{equation}
and is related to the eigenvalue $c_2(R)$ of the quadratic Casimir operator,
\begin{equation}
	c_2(R) = \delta^{AB}T_R^A T_R^B,
\end{equation}
in the representation $R$ by
\begin{equation}
	t_2(R) = \frac{\dim R}{\dim \mathfrak{su}(3)} c_2(R),
\end{equation}
where $\dim R$ and $\dim \mathfrak{su}(3)=8$ denote the dimensions of the representation and the Lie algebra $\mathfrak{su}(3)$ of the gauge group $\text{SU}(3)$, respectively. Labelling irreducible representations $R(p,q)$ of $\mathfrak{su}(3)$ by the Dynkin label $(p,q)$, the eigenvalue of the quadratic Casimir is
\begin{equation}
	c_2(R(p,q)) \equiv c_2(p,q) = \frac{p^2+q^2+3p+3q+pq}{3} \, .
\end{equation}
Moreover, the dimension of the representation $R(p,q)$ can be calculated by
\begin{equation}
	\dim R(p,q) = \frac{(p+1)(q+1)(p+q+2)}{2} \, .
\end{equation}

In the Standard Model, gluons transform with respect to the adjoint representation $\boldsymbol{8}$ and fermions with respect to the fundamental representation $\boldsymbol{3}$. Therefore, with $\nf=6$, $\NC=3$, and $\TR=1/2$, in the Standard Model we have
\begin{equation}
	t_2(V) = t_2(\boldsymbol{8}) = 3, \quad \text{and} \quad t_2(F) = t_2(\boldsymbol{3}) = \frac{1}{2}.
\end{equation}

Some Clebsch-Gordan decompositions useful to determine the terms entering the scalar potential are tabulated below. All decompositions are calculated with \texttt{LieART} \cite{Feger:2019tvk} and cross-checked with an independent implementation of the algorithm in \cite{Coleman:1964su}.

\begin{table}[h]
	\centering
	\caption{Selected three- and four-particle Clebsch-Gordan decompositions.}
	\begin{tabular}{cl}\toprule
		\textbf{Direct Product} & \textbf{Decomposition} \\ \midrule
		$\mathbf{3} \otimes \mathbf{3} \otimes \mathbf{3}$ &  $\mathbf{1} \oplus_2 \mathbf{8} \oplus  \mathbf{10}$  \\ 
		$\mathbf{6} \otimes \mathbf{6} \otimes \mathbf{6}$ &  $\mathbf{1} \oplus_2 \mathbf{8} \oplus  \mathbf{10} \oplus \overline{\mathbf{10}} \oplus_3  \mathbf{27} \oplus  \mathbf{28} \oplus_2 \mathbf{35}$\\ 
		$\mathbf{8} \otimes \mathbf{8} \otimes \mathbf{8}$ &  		
		$\oplus_2 \mathbf{1} \oplus_8 \mathbf{8}  \oplus_4 \mathbf{10} \oplus_4 \overline{\mathbf{10}}  \oplus_6 \mathbf{27} \oplus_2 \mathbf{35} \oplus_2 \overline{\mathbf{35}}\oplus \mathbf{64} $  \\ 			
		$\mathbf{10} \otimes \mathbf{10} \otimes \mathbf{10}$ &  		
		$ \mathbf{1} \oplus_2 \mathbf{8}  \oplus \mathbf{10} \oplus \overline{\mathbf{10}}  \oplus_3 \mathbf{27} \oplus \mathbf{28} \oplus_2 \mathbf{35} \oplus_2 \overline{\mathbf{35}} \oplus \cdots $  \\ 
		$\mathbf{15} \otimes \mathbf{15} \otimes \mathbf{15}$ &  		
		$ \oplus_2\mathbf{1} \oplus_{10} \mathbf{8}  \oplus_8 \mathbf{10} \oplus_7 \overline{\mathbf{10}}  \oplus_{15} \mathbf{27} \oplus_4 \mathbf{28} \oplus \overline{\mathbf{28}}\oplus_{11} \mathbf{35} \oplus_8 \overline{\mathbf{35}} \oplus \cdots $  \\ 
		$\mathbf{15^\prime} \otimes \mathbf{15^\prime} \otimes \mathbf{15^\prime}$ &  		
		$ \mathbf{1} \oplus_{2} \mathbf{8}  \oplus \mathbf{10} \oplus \overline{\mathbf{10}}  \oplus_{3} \mathbf{27} \oplus \mathbf{28} \oplus \overline{\mathbf{28}}\oplus_{2} \mathbf{35} \oplus_2 \overline{\mathbf{35}} \oplus \cdots $  \\ \midrule							
		$\mathbf{8} \otimes \mathbf{8} \otimes \mathbf{8} \otimes \mathbf{8} $ & 	
		$\oplus_8 \mathbf{1} \oplus_{32} \mathbf{8}   \oplus_{20} \mathbf{10} \oplus_{20} \overline{\mathbf{10}} \oplus_{33} \mathbf{27} \oplus_2 \mathbf{28} \oplus_2 \overline{\mathbf{28}}  \oplus_{15} \mathbf{35} \oplus_{15} \overline{\mathbf{35}} \oplus \cdots$ \\
		$\mathbf{10} \otimes \mathbf{10} \otimes \mathbf{10} \otimes \mathbf{10} $ & 
		$\mathbf{1}\oplus_{8} \mathbf{8}  \oplus_{10} \mathbf{10} \oplus_{6} \overline{\mathbf{10}} \oplus_{15} \mathbf{27}  \oplus_6 \mathbf{28} \oplus_{4} \overline{\mathbf{28}}   \oplus \cdots$
		 \\
		$\mathbf{10} \otimes \mathbf{10} \otimes \mathbf{10} \otimes \overline{\mathbf{10} }$ & 
		$\mathbf{1}\oplus_{8} \mathbf{8}  \oplus_{7} \mathbf{10} \oplus_{10} \overline{\mathbf{10}} \oplus_{18} \mathbf{27}  \oplus_{10} \mathbf{28} \oplus_{3} \overline{\mathbf{28}}   \oplus \cdots$
		 \\
		\bottomrule
	\end{tabular}
	\label{prods34}
\end{table}

\begin{table}[h]
	\centering
	\caption{Relevant Clebsch-Gordan decompositions needed to form a color-singlet bound state with one colored scalar.}
	\begin{tabular}{cl}\toprule
		\textbf{Direct Product} & \textbf{Decomposition} \\ \midrule
		$\mathbf{3} \otimes \overline{\mathbf{3}} $ & $\mathbf{1}\oplus \mathbf{8} $ \\
		$\mathbf{8} \otimes \mathbf{8}$ & $\mathbf{1}\oplus _2\mathbf{8} \oplus \mathbf{10} \oplus \overline{\mathbf{10}}  \oplus \mathbf{27}$ \\
		$\mathbf{3} \otimes {\mathbf{3}}  \otimes {\mathbf{3}} $ & $\mathbf{1}\oplus _2\mathbf{8} \oplus \mathbf{10} $\\
		$\mathbf{3} \otimes {\mathbf{3}}  \otimes {\mathbf{8}} $ & $\mathbf{1}\oplus _3\mathbf{8} \oplus \mathbf{10} \oplus \overline{\mathbf{10}}  \oplus \mathbf{27}$\\
		${\mathbf{3}} \otimes {\mathbf{6}}  \otimes \mathbf{8} $ & $\mathbf{1}\oplus_3\mathbf{8} \oplus_2 \mathbf{10} \oplus \overline{\mathbf{10}}  \oplus_2 \mathbf{27} \oplus \mathbf{35}$\\
		$\overline{\mathbf{3}} \otimes \overline{\mathbf{3}}  \otimes \mathbf{6} $ & $\mathbf{1}\oplus _2\mathbf{8} \oplus \mathbf{10}  \oplus \mathbf{27}$\\
		$\overline{\mathbf{3}} \otimes {\mathbf{8}}  \otimes \mathbf{15} $ & $\mathbf{1}\oplus_4\mathbf{8} \oplus_3 \mathbf{10} \oplus_2 \overline{\mathbf{10}}  \oplus_4 \mathbf{27} \oplus \cdots$\\
		${\mathbf{8}}  \otimes \mathbf{8} \otimes {\mathbf{10}}$ & $\mathbf{1}\oplus_4\mathbf{8} \oplus_4 \mathbf{10} \oplus_2 \overline{\mathbf{10}}  \oplus_5 \mathbf{27} \oplus \cdots$\\
		$\overline{\mathbf{3}} \otimes {\mathbf{8}} \otimes {\mathbf{8}} \otimes \mathbf{15}^\prime $ & $\mathbf{1}\oplus_6\mathbf{8} \oplus_8 \mathbf{10} \oplus_3 \overline{\mathbf{10}}  \oplus_{11} \mathbf{27} \oplus \cdots$\\
		\bottomrule
	\end{tabular}
\label{tab:rhadronsCG}
\end{table}

\section{Scalar loop functions}\label{sec:loopfun}
In this appendix, we collect the well-known scalar one-loop functions appearing in the loop-induced diagrams $H\to gg$ and $S \to gg$.
\begin{eqnarray}
I_q(x) &=& 2x - x\ (1-4x) f(x),~~~ I_s(x) = -(1+2x f(x)), \nonumber \\
f(x)&=&\left\lbrace
\begin{array}{ccc}
\dfrac{1}{2}\left(\ln\left(\dfrac{1+\sqrt{1-4x}}{1-\sqrt{1-4x}}\right)-i\pi\right)^2 & & \text{for }x<\dfrac{1}{4}\\ & & \\
-2\left(\arcsin\left(\dfrac{1}{2\sqrt{x}}\right)\right)^2 & & \text{for }x>\dfrac{1}{4}
\end{array}\right.
\label{loopf}
\end{eqnarray}

\bibliography{refs.bib}

\end{document}